\documentclass[preprint2]{aastex}
\doublespace
\onecolumn

\newcommand{\etal}{{et~al.~}}

\newcommand{\cc}{\mbox{${\rm cm}^{-3}$}}

\newcommand{\kms}{\mbox{${\rm km~s}^{-1}$}}

\newcommand{\vlsr}{\mbox{$\rm V_{\rm LSR}$~}}

\newcommand{\htwo}{\mbox{${\rm H}_2$~}}

\slugcomment{Submitted to The Astrophysical Journal}

\shortauthors{Brunt \& Heyer}
\shorttitle{Diagnostics to Interstellar Turbulence}

\begin{document}

\def\gtabouteq{\,\hbox{\raise 0.5 ex \hbox{$>$}\kern-.77em
                    \lower 0.5 ex \hbox{$\sim$}$\,$}}
\def\ltabouteq{\,\hbox{\raise 0.5 ex \hbox{$<$}\kern-.77em
                     \lower 0.5 ex \hbox{$\sim$}$\,$}}
\def\vlsr{V$_{LSR}$}
\def\kms{km s$^{-1}$}

\title{Interstellar Turbulence:\\
     II. Energy Spectra of Molecular Regions in the Outer Galaxy }

\author{Christopher M. Brunt}
\affil{Five College Radio Astronomy Observatory, Department of 
Astronomy, Lederle Research Building,
University of Massachusetts, Amherst, MA 01003, USA}
\affil{National Research Council, Herzberg Institute of Astrophysics, Dominion Radio Astrophysical Observatory, Penticton, BC, CANADA}
\affil{Department of Physics and Astronomy, University of Calgary, CANADA}
\and
\author{Mark H. Heyer}
\affil{Five College Radio Astronomy Observatory, Department of 
Astronomy, Lederle Research Building,
University of Massachusetts, Amherst, MA 01003, USA}

\begin{abstract}
The multivariate tool of Principal Component Analysis (PCA) is applied 
to 23 fields in the FCRAO CO Survey of the Outer Galaxy. 
PCA enables the identification of line profile 
differences which are assumed to be generated from fluctuations within a
turbulent velocity field.   The variation of these velocity differences
with spatial scale within a molecular region is described by a singular power
law, ${\delta}v=cL^\alpha$ which can be used as 
a powerful diagnostic to turbulent motions.   For the ensemble of 23 fields,
we find a mean value $<\alpha>$=0.62$\pm$0.11. 
From a recent calibration of this method using fractal Brownian motion 
simulations (Brunt \& Heyer 2001), the measured 
velocity difference-size relationship corresponds to an energy spectrum,
$E(k)$, which varies as k$^{-\beta}$, where $\beta=2.17\pm0.31$. 
 We compare our results to 
both decaying and forced hydrodynamic simulations of turbulence. 
We conclude that energy must be continually injected into the regions
to replenish that lost by dissipative processes such as shocks.
The absence of large, widely distributed shocks within the targeted fields suggests 
that the energy is injected at spatial scales less than several pc.
\end{abstract}
\keywords{hydrodynamics --- turbulence --- 
ISM: kinematics and dynamics --- ISM: clouds
line: profiles --- 
methods: statistical}

\section{Introduction}

Complex velocity fields and non-thermal line widths attest to the presence of 
turbulent flows in dense, molecular clouds.   The energy density 
associated with such flows may provide the 
overall dynamical support to counter the self-gravity of molecular clouds and
the pressure of the external medium and therefore, regulates the 
formation of stars in the Galaxy (Zuckerman \& Evans 1974).  
For compressible gas, the advection of material through a chaotic 
velocity field contributes to the hierarchical structure of the 
dense interstellar medium (Falgarone, Phillips, \& Walker 1991).  
While such dynamical functions have long been 
assigned to turbulence, there are few observational descriptions
of the physical nature of such flows.  Such descriptions are 
essential to a more complete accounting of the star formation process.

Since the instantaneous behavior
of turbulent flows is unpredictable,
most theoretical descriptions of turbulence 
focus upon statistical
correlations of the velocity field (Landau and Lifshitz 1959). 
The energy spectrum, E(k), defined as the angular integral of the 
power spectrum 
of the velocity field, quantifies the degree to which the velocities
are correlated with spatial scale.   
Over the inertial range of spatial scales, 
the energy spectrum varies as a singular power law
with wave vector k,
$$ E(k) \sim k^{-\beta} \eqno(1.1) $$
The scaling exponent, $\beta$, provides a concise description of the 
velocity field and can be used to distinguish between different 
turbulent conditions.  For example, the 
incompressible energy cascade model of Kolmogorov
(1941) gives $\beta = 5/3$.  For compressible 
turbulent flows in which energy is 
dissipated in ubiquitous shocks, $\beta=2$ (Passot, Pouquet, \& Woodward 1988;
Passot, Vazquez-Semadeni, \& Pouquet 1995; Gammie \& Ostriker 1996). 
The functional form of the energy spectrum in 
equation 1.1 is equivalent to the statement that the 
root mean square velocity varies with size $l$ over which it is measured as
$$ <v_l^2>^{1/2} \sim l^\gamma \eqno (1.2) $$
where $\gamma=(\beta-1)/2$.

In principle, values of $\beta$ applicable to the molecular interstellar 
medium are constrained by observations.  In practice, 
retrieving such dynamical information from spectroscopic imaging
observations is challenging.  
Observed radiation fields lie
on a fundamentally different basis than the intrinsic
physical fields. Both real and modeled interstellar clouds 
are described by 
physical fields of velocity,
v(x,y,z), density, n(x,y,z),
temperature, T(x,y,z), molecular abundance, and the local UV radiation field,
where x,y,z are {\it spatial} 
coordinates.
Observed radiation fields, T$_{R}$(x,y,v$_{z}$), of a given 
molecular line transition
are retrieved on the basis of projected position on the
sky (x,y) and the Doppler-shifted line of sight velocity
(v$_{z}$). Moreover, the radiation temperature field
is an extremely complicated convolution of velocity, density,
temperature and optical depth structure along the line
of sight. 
If the velocity field is turbulent,
involving multiple non-local re-correlations along the
line of sight, then
the transformation of the field to an axis is highly non-linear.

Several studies have shown that the observed molecular line profiles
are incompatible with extreme values of $\beta$ (Goldreich \& Kwan
1974; Leung \& Lizst 1976; Dickman 1985; Kwan \& Sanders 1986).
That is, the velocity field is neither smooth ($\beta > 3$) nor
micro-turbulent ($\beta < 1$).  More direct measures of $\beta$
include
``cloud-finding''
decompositions from which a size-line width relationship is 
identified (Carr 1987; St\"utzki \& G\"usten 1990) 
and autocorrelation function and structure
function analyses of centroid velocity fluctuations (Scalo 1984; 
Kleiner \& Dickman 1985; Miesch \& Bally 1994).
While these analyses are widely used, there has 
been no demonstration that these methods can
accurately recover the statistics, (i.e. $\beta$ or equivalently, $\gamma$), of
molecular cloud velocity fields from radiation
temperature (T$_{R}$(x,y,v$_{z}$)) measurements
over a wide range of conditions.  More recent efforts to exploit the 
three dimensional information within the observed data cubes include 
Rosolowsky \etal (1999) and Lazarian
\& Pogosyan (2000).   

Heyer and Schloerb (1997)
have applied the multivariate 
technique of Principal Component Analysis (PCA) 
to decompose spectroscopic imaging observations
of targeted molecular regions.  The effect of the decomposition is to 
identify 
velocity gradients 
over various scales as these are measured by {\it differences within the line 
profiles}
with respect to the noise and resolution limits.  
From the set of significant eigenvector and eigenimages, they derive 
a power law relationship between  the velocity differences 
and the spatial scale over which
these gradients occur such 
that
$$ {\delta}v \sim L^\alpha \eqno (1.3) $$
In a companion paper, Brunt \& Heyer (2001) demonstrate
that the method outlined by Heyer \& Schloerb (1997) with some
modifications,  is sensitive to 
velocity fields with varying statistics and provide a calibration to estimate
$\beta$ from the observed power law index $\alpha$
under a wide variety of physical and observing 
conditions.  The basic calibration is 
$\beta = 3{\alpha}+0.15 $. We stress that
this calibration was obtained under highly
idealized conditions, using simulated
observations generated from fractional
Brownian motion (fBm) velocity fields and
multifractal density fields, as described
in Brunt \& Heyer (2001). While such a calibration allows
statistically controlled inputs to the
calibration, it does not include features
expected in real turbulent fields, such
as intermittency or correlations between
velocity convergences and density
enhancements. For observational
concerns, Brunt \& Heyer (2001) found that
PCA was not strongly affected by
emission saturation, which may be expected
to be present in the $^{12}$CO (J=1--0)
data utilized here. The reasons for
this are not completely understood, since 
the PCA method is currently empirical in
nature. Self-reversal, we expect, would
have an impact, but this was not 
prevalent in the simulated observations,
nor is it widely seen in $^{12}$CO (J=1--0)
emission. These points are
discussed in more detail in $\S$4.

In this paper, we apply the PCA method to an ensemble of molecular regions
to determine the degree of velocity correlation within the interstellar
medium.  The fields are 
extracted from the FCRAO CO Survey of the Outer Galaxy (Heyer \etal 1998).
In $\S$2, we describe how the regions were selected.  The analysis
method is demonstrated in detail for one of the targeted regions in $\S$3.
A summary of the analysis upon all of the fields is presented in 
$\S$4.  In $\S$5, we discuss the results in context 
with phenomenological 
descriptions and hydrodynamic simulations of
interstellar turbulence.
\section{The Data}

The FCRAO CO Survey of the Outer Galaxy is a spectroscopic imaging survey of
molecular $^{12}$CO emission
over a very large region of the outer Galaxy
($40^\circ\times8^\circ$, sampled every $50\arcsec$)
and covering the \vlsr~ range -150 to +40 \kms~
sampled every 0.81 \kms~ (Heyer \etal 1998).
The wide field and spatial dynamic range provide 
an unbiased sample of molecular emission
and enable the isolation of a large
number of fields to analyze.

\subsection{Selected Fields}

Twenty three fields have been selected from the Outer Galaxy Survey
with a broad range of cloud morphologies.
Nine of the fields are within the 
Perseus
spiral arm defined here spectroscopically as 
$-60 < V_{LSR} < -30$ \kms.  
The remaining fields have centroid velocities greater than 
-20 \kms~ and are 
referred to
collectively as ``local'' fields. 
Several well-known
giant molecular clouds, both in the Perseus and local arms 
are included in the sample.  These include clouds associated
with optical and radio HII regions and OB associations (W3, W5, NGC~7538, and 
Sh~156, and Cep~OB3). 
Fields were visually selected from
inspection of spectroscopically-restricted
integrated intensity maps and spatially-restricted
global spectrum plots. 
In some cases, due to the angular limits of the 
Survey or due
to the imposition of spectroscopic segregation, the
emission does not fall to the zero level at the 
boundaries for some fields. This feature is mostly restricted to the
local emission, which covers a broader range in 
Galactic latitude and suffers more from 
spectroscopic blending.  
Tests have been carried
out to gauge the effect of truncating a region and suggest that it does
not cause any significant problems since we are primarily interested
in the statistics of the fields which are well sampled by the available 
spectra.

An overview of the selected fields is presented in
Figure~1.
The spectroscopic and angular limits ($v_{min}$:$v_{max}$, 
$l_{min}$:$l_{max}$, $b_{min}$:$b_{max}$),
the number of spectroscopic channels (N$_{v}$)
and spatial pixels (N$_{l}$ $\times$ N$_{b}$)
of these fields are reported
in Table~1.
To identify these fields in Figure~1, 
the identification number of the
Perseus Arm (P) and local (L) fields increases  
with decreasing $l_{max}$. 
To obtain absolute spatial scales, kinematic distances have
been derived 
from the velocity and positional centroids
in each field
assuming a flat rotation curve
with V$_\circ$ = 220 kms$^{-1}$ and 
R$_{\circ}$ = 8.5 kpc.
The derived scaling exponent, $\alpha$, is
independent of distance while the amplitude, ${\delta}v$(1pc), is 
linearly dependent upon the presumed distance to the source.
In this study, we are exclusively interested in $\alpha$.

\section {Results}
\subsection{A Detailed Example of the Analysis}

The formal description of PCA is presented by Heyer \& Schloerb (1997)
 and Brunt \& Heyer (2001).
Here, we provide a detailed example of the complete analysis upon 
an individual field to
demonstrate the statistical output from PCA and the 
extraction of characteristic scales from the 
set of eigenimages and eigenvectors. 
Given a data set $T(x_i,y_i,v_j)=T_{ij}$ 
with n=$N_l{\times}N_b$ spectra and P spectroscopic channels,
a set of P eigenvectors, $u_l$, and eigenvalues, $\lambda_l$,
are derived from 
the solution of the eigenvalue equation with the 
covariance matrix, S, of the data, 
$$ Su = \lambda u \eqno(3.1)$$
where
$$ S_{jk} = {{1}\over{n}} \sum_{i=1}^n T_{ij} T_{ik} \eqno(3.2) $$
The eigenvectors are subject to the orthogonal condition, 
$$
\sum_{j=1}^{P} u_{jl} u_{jm} = \left\{
\begin{array}{ll}
  1 & \mbox{ if  $l=m$} \\
  0 & \mbox{ if $l{\neq}m$}
\end{array}
\right \}
$$
The eigenvalues are equivalent  to the variance of the data projected 
onto the respective eigenvectors.
The eigenvalues and corresponding eigenvectors are sorted from
largest to smallest to establish the principal components of the data set.
An eigenimage, $I_l$, is constructed by projecting each spectrum in the field
onto the $l^{th}$ eigenvector, $u_l$,
$$ I_l(x_i,y_i)=I_l(r_i) = \sum_{j=1}^{P} T_{ij} u_{jl}  \eqno(3.4) $$
Effectively, a given eigenvector, $u_l$, 
defines a spectral window with a characteristic
velocity, ${\delta}v_l$, which gauges the profile differences
over the observed field.  The spectra are weighted by $u_l$ 
to produce a spatial map, $I_l(r)$, which locates line profile differences
within the data cube. 
%% A convenient and important property of the 
%% decomposition is that the 
%% variance of the projected noise values is equal to that of the noise of the 
%% data.
The first principal component generally consolidates
all of the spectroscopic channels with signal.  Higher order 
components ($ l \approx P$) provide an 
accurate measure of the variance due to the 
random noise of the data.  The equivalence of the eigenvalues with the 
projected variance enables a simple characterization of the signal to noise
ratio of the data cube, 
$$ \zeta_1 = \sqrt{ {\lambda_1 - \lambda_{P}}\over {\lambda_{P}} } \eqno(3.5) $$

In Figure~2 and Figure~3, the set of eigenvectors and eigenimages 
derived for field P6 (NGC~7538) are shown.  The first 3 components ($l=1,2,3$)
identify the large scale kinematics of the cloud which is comprised of 
2 features at velocities -56 and -50 \kms~ respectively. 
Subsequent components ($l=4,5,6$) isolate smaller velocity 
differences within these features as these contribute 
to the measured variance of the data cube.  
Finally, for $l\ge8$, any variance
due to cloud kinematics can not be distinguished from the random noise
of the data.   Therefore, for this cloud,  we need only to consider components $l<8$.

The identification of smaller velocity differences by the eigenvectors 
with increasing $l$ 
is intrinsic to the use of orthogonal functions to decompose the data cube.
However, the spatial granularity of the eigenimages also decreases with 
increasing $l$ and reflects a fundamental property of spectral line 
observations of the molecular gas component in which line profiles are 
increasingly similar in shape and velocity with decreasing angular distances.
It is this variation of spatial granularity with decreasing velocity
which we wish to
quantify as this is related to the energy spectrum of the velocity field 
(Brunt \& Heyer 2001).

To determine the characteristic velocity and spatial scales for each
component, the raw autocorrelation functions, $C_V^l(dv)$, $C_I^l(\tau)$, 
are calculated from the 
eigenvectors and eigenimages respectively,
$$ C_{V}^{l}(dv) = < (u^{l}(v) u^{l}(v+dv)> \eqno(3.6a) $$
$$ C_{I}^{l}(\tau) = < (I^{l}(r) I^{l}(r+\tau)> \eqno(3.6b) $$
where $\tau=\tau(x,y)$
These expressions include an additive component, $C_N(\tau)$ due to 
instrumental noise (Brunt \& Heyer 2001). 
 The noise subtracted autocorrelation function,
$C_{I0}(\tau)$ is calculated,
$$  C_{I0}(\tau) = C_I(\tau) - C_N(\tau) \eqno(3.7) $$
Since the 
2 dimensional autocorrelation function is derived from a large number of lags,
the noise contribution can be significant.  Moreover, for spectroscopic data
cubes constructed using reference sharing or on the fly mapping modes,
the noise is correlated between positions observed with the same receiver
horn and reference position.  In these cases, $C_N(\tau) \neq 0$ 
for $\tau \neq 0$.
The details to deriving $C_N(\tau)$ 
are discussed in the 
Appendix.
The noise subtracted autocorrelation functions
determined from the set of eigenvectors
and eigenimages displayed in Figure~2 and Figure~3 are shown
in Figure~4 and Figure~5 respectively.

The characteristic velocity scale, ${\delta}v_l$, is determined
from the velocity lag at which 
$$C_V^l({\delta}v_l)/C_V^l(0) = e^{-1} \eqno(3.8) $$
For the 2 dimensional ACF of the eigenimage, one must additionally 
correct for the 
effects of finite resolution observations upon the zero lag (see Brunt \& Heyer 2001).
We determine the spatial
correlation lengths along the cardinal directions, ${\delta}x_l, {\delta}y_l$,
 from the noise corrected
ACF,
$$ { {C_{I0}^l({\delta}x_l,0)}\over{C_{I0}^l(0,0)} } =e^{-1} \eqno(3.9a) $$
$$ { {C_{I0}^l(0,{\delta}y_l)}\over{C_{I0}^l}(0,0) } =e^{-1} \eqno(3.9b) $$
and assign the {\it biased} characteristic spatial scale, $L_{lB}$ to the
quadrature sum,
$$ L_{lB} = \sqrt{{\delta}x_l^2+{\delta}y_l^2} \eqno(3.10) $$
The true characteristic scale, $L_l$, 
is derived from $L_{lB}$ with subpixel corrections to account for 
finite resolution of the observations.
Since the eigenimages tend to be isotropic, the ACF profiles along 
the cardinal directions are a reasonable approximation to the variation 
of the ACF along any arbitrary angle.
The characteristic velocity and spatial scales are statistically well defined 
with respect to the noise as these represent mean values over the full spectrum
and field respectively.   A measurement error, $\sigma_{{\delta}v}$, to the
value ${\delta}v_l$ is given by one half the velocity resolution
of the spectrometer.  For the spatial scale, $\sigma_L$ is determined from
the quadrature sum of the spatial resolution 
and the degree of 
anisotropy, $|{\delta}x_l-{\delta}y_l|$, in the spatial ACF.
The PCA velocity statistic, $\alpha$, is then 
obtained from the set of ${\delta}v,L$ pairs which are larger than the 
respective spectral and pixel resolution limits and form the power law 
relationship
$$ {\delta}v \propto L^{\alpha} $$
where the unsubscripted quantities refer to the ensemble
(${\delta}v_l, L_{l}$). The retrieval
of each ${\delta}v,L$ pair
represents a vast amount
of information consolidated from the entire field
and describes the spatial scales over which velocity gradients are
detected as line profile differences.
The set of (${\delta}v$, L) points derived from the 
significant principal components are plotted in Figure~6.
A power law fit to these points yields a scaling exponent $\alpha=0.69\pm0.02$.

\subsection{Results for All Targeted Fields}

Individual plots of the retrieved characteristic scale
measurements for the selected fields are shown
in Figure~6 and Figure~7.
Power law fits to these relationships yield the scaling exponent, $\alpha$ 
and the amplitude, $c={\delta}v(1pc)$.  These values 
are listed in Table~2 in addition to 
the number of principal components included in the fit (N$_{pc}$),
the number of principal components rejected (N$_{c}$ - see Brunt 1999),
and the variance-based signal-to-noise estimates ($\zeta_{1}$).
Values of $\gamma$ and $\beta$  which describe the 
intrinsic velocity field statistics have been 
derived according to the calibration of Brunt \& Heyer (2001) and are also 
tabulated in Table~2.
The error estimates listed for these values include all fitting errors and
calibration uncertainties. 

A summary of the derived values for $\alpha$ and $\beta$ is presented
as a histogram in Figure~8.  The shaded part of the histogram 
includes those regions with associated HII regions and OB stars.
We have calculated weighted (1/$\sigma_\alpha^2$, where $\sigma_\alpha$ 
is the statistical
uncertainty in the scaling exponent)
and unweighted averages of
the exponents reported in Table~2.
The weighted values are: $<\alpha>=0.62\pm0.11$, $<\beta>=2.17\pm0.31$ ,
$<\gamma>=0.49\pm0.16$.  The unweighted values are:
$<\alpha>=0.68\pm0.10$,
$<\beta>=2.21\pm0.31$, $<\gamma>=0.57\pm0.13$. The 
uncertainties of these derived averages are larger 
than those of the individual fits. 
The range in $\gamma$
is comparable to the range found by centroid velocity
ACF analysis (see Miesch and Bally 1994).

There is no general relationship between
the measured value of $\alpha$ and any subjectively assigned
morphology of the fields or association with
HII regions and OB stars. 
It is also evident that the PCA results for
the local fields as a whole contain a larger
degree of scatter than the Perseus Arm
fields. There is no reason of course to
expect that these relationships should be
exact power laws, and the instances of large
scatter may be indicative of complex systematic
motions or multi-component field statistics which 
are not resolved by our analysis.

\subsection{ Large-Scale Field Results}

The angular and spectroscopic boundaries of the targeted fields 
have been determined by simple
inspection of integrated images over limited velocity intervals.
As such, some fields are truncated at the boundary 
or may be comprised of two or more unrelated clouds which share a common
velocity.   If two or more fields with varying velocity fields statistics
are grouped together as a singular field, this inhomogeneity may 
skew the estimate to $\alpha$. 
To gauge the effect of inhomogeneous velocity fields upon the composite field, 
we have constructed a
large scale field which contains the selected fields
P4, P6, P7, P8 and P9.  PCA is applied to this field and 
scale measurements are determined. 
The results are shown in Figure~9.
The solid line is the fit, $\alpha = 0.68$, c = 1.47 kms$^{-1}$.
The derived exponent is close to the mean value of the 
individual fields (0.65).  
The largest characteristic scales obtained from this
measurement are very similar to those obtained from the
individual fields. 

Secondary considerations are the effects of large scale flows such 
as Galactic rotation or streaming motions from a spiral potential.
The rotation of the Galaxy generates a 
velocity difference across the field,  
$${\Delta}v=({\partial}v_R/{\partial}l){\Delta}l
            \eqno(3.11)$$
$${\Delta}v
           = cos(l)(V(R)(R_\circ/R)-V(R_\circ)){\Delta}l \eqno(3.12)$$
where R is the galactocentric radius of the cloud, V(R) is the 
rotation curve, and ${\Delta}l$ is the longitude extent of the field 
in radians.
A signature of Galactic rotation would be 
a secondary, linear component at the largest scales in the measured power law.  
For molecular regions in this study, the angular extents are 
small (${\Delta}l < 3^\circ$).  Assuming a flat rotation curve
($V(R)=V(R_\circ)$=220 \kms),
any velocity shifts due to Galactic rotation are small (${\Delta}v < 2$ \kms)
or comparable to the velocity resolution of the data (0.98 \kms). 
Application of this method to data cubes with signal 
more continuously 
distributed over large scales, such as the HI 21cm line,  would more 
readily detect this signature of Galactic rotation.

\section{Discussion}

The analysis used in this study 
identifies gradients within the data cube with respect to 
the noise level and resolution limits, as these gradients are reflected 
in line profile differences.  
These are direct statistical measurements 
with no implicit assumptions or
predefined descriptions of clumps within the cloud complex.  
Such differences arise from various
dynamical processes such as rotation, collapse, outflow, 
turbulence and on larger scales, streaming due to spiral arm
potentials or rotation of the Galaxy.  
Systematic collapse motions over the entire cloud 
are unlikely.  Similarly, velocity gradients due to rotation 
across the 
projected surface of a cloud 
are rarely identified.
In general,  systematic, smooth velocity fields 
generate values of $\beta \ge 3$ and are inconsistent with those 
values measured in Figure~8.   As was shown in $\S$3.3, the observed 
motions are larger than those due to the rotation of the Galaxy.
Thus, we conclude that the 
identified gradients are primarily due to fluctuations
within the turbulent velocity fields of the targeted regions.

The relationship between the measured
scaling exponent, $\alpha$, and the exponents which describe the 
velocity field, ($\beta$ or $\gamma$) is derived from fBm simulations
(Brunt \& Heyer 2001).  The fBm model clouds, while statistically 
well defined, are physically unrealistic in that the velocity and 
density fields are uncorrelated.  In reality one would expect 
these fields to be highly interdependent in a compressible medium due to 
the advection term in the fluid equations which govern the time 
evolution of material.  Also, in the fBm models, 
the velocity field is equally 
distributed between solenoidal and longitudinal modes.  Hydrodynamic 
simulations often generate velocity fields which are largely solenoidal.
Finally, the fBm simulations do not account for intermittency in which there 
are strong, non-gaussian velocity fluctuations although it is not 
evident whether PCA is even sensitive to such effects.  

Finally, we consider the possibility of systematic
biases in our measured exponents, aside from the
caveats listed above. As shown in Paper I,
PCA is mostly unaffected by saturation of the emission, as
is certain to occur for the $^{12}$CO (J=1-0) line. Indeed,
the more highly saturated measurements are
more reliable than less saturated measurements, due mainly
to the fact that more of the velocity field is sampled
and hence a fuller account of the velocity field
statistics is obtained.
Variance in the observed line profiles under saturated
conditions is still observed, arising from variation in
the line centroid velocities. In the saturated regime,
PCA preferentially traces variance due to
this line centroid variation, while under less saturated
conditions PCA works more similarly to 'cloud-finding'
analyses (see Heyer \& Schloerb 1997), by identifying
kinematically distinct structures. 

While PCA provides an effective bridge between
these two regimes, it is not entirely without bias
over the full range of possibilities. In Paper I,
we identify regions of low intrinsic $\beta$ ($\sim$ 1)
which are sparsely sampled by the molecular tracer (eg
as in a C$^{18}$O observation) as being more
sensitive to bias. Specifically, an inferred intrinsic
$\beta$ from an observation in this regime is
prone to overestimation.
For our most sparsely sampled observations at
$\beta$ = 1, we derive $\beta$ = 1.35 - 1.56
($\alpha$ = 0.4 - 0.47, instead of the
expected $\alpha \sim $ 0.3), while
higher values of intrinsic $\beta$ are not biased. 

The origin of this bias comes from the characteristic
scale measurements being skewed preferentially towards
the {\it typical size} of the isolated kinematic structures
rather than their typical separation, in accord
with our description of PCA under conditions of
sparse sampling and our method of deriving the
characteristic spatial scales through ACF analysis.
Our observations here are not sparsely sampled and
so do not suffer this bias strongly, nor are the
recovered values of $\alpha$ consistent with such
overestimation since they lie in the regime 
($\beta \geq$ 2 generally) which is not subject to
this sparse sampling bias.

We do note however that accurate characterization
of regions of intrinsically low $\beta$ which
are sparsely sampled by the molecular tracer would
not be possible with our method in
its current formulation. This shortcoming may
be possibly circumvented in the future by
a modified method by which characteristic scales
are measured or by estimating potential bias
through analysis of the integrated
intensity power spectrum, as can be done in the 
power spectrum method of Lazarian \& Pogosyan (2000)
for which  similar type of sampling bias occurs.

\subsection{Comparison with Numerical Models of Interstellar Turbulence}

Magnetohydrodynamic simulations 
provide  physical insight to the complex, turbulent flows 
within the interstellar 
medium under a variety of conditions.   However, given the 
strong dependence on initial conditions, the simulations require
established observational constraints in order to 
refine the input parameters and to gauge the physical 
relevance of the numerical 
results.
Ideally, the measured distribution of $\beta$ shown in Figure~8
provides a fundamental 
constraint to E(k) generated by 
the increasingly more sophisticated 
hydrodynamic simulations and phenomenological models of interstellar 
turbulence (Ostriker, Gammie, \& Stone 1999; Padoan, Jones, \& Nordlund 1997;
Vazquez-Semadeni, Passot, \& Pouquet 1995).  
In practice,  direct comparisons do not account for the filtered 
nature of molecular line observations.  
That is, the velocity field can only be probed 
in those regions in which the chosen molecular tracer is detected. 
In low column density regions, 
the ambient 
ultraviolet radiation field 
can modify the gas phase by photodissociation.  In regions of low
volume density (n(\htwo) $<$ 50 \cc), the molecule may not be 
sufficiently excited by collisions or radiative trapping 
to be detected.  The effect
of these conditions is to mask a portion of the velocity field and 
limit the range of wave vectors traced
by the data.
In contrast, the 
energy spectrum for a simulation is calculated directly 
from the full velocity field with no weighting of the density 
field or consideration of radiative transfer effects.  
Therefore, a more relevant and fair comparison would be to construct an 
ideal observation, $T_A(x,y,V)$,
from the velocity, temperature, and density fields (Falgarone \etal 
1994; Padoan \etal 1998; Brunt \& Heyer 2001) 
and to apply the same
statistical analyses described in $\S$3.  
Such ideal observations are not presently available so we 
make direct comparisons between the data and recent 
simulations while considering the above limitations.

The Kolmogorov spectrum 
($\beta=5/3$) describes the energy spectrum of an 
incompressible fluid in which energy is deposited at the largest scales
and cascades to the smallest scales where it is ultimately 
dissipated (Kolmogorov 1941).  Within the inertial range, 
the kinetic energy resides within 
solenoidal (divergenceless) motions or eddies.  
While this provides an idealized
description of an incompressible turbulent flow, several 
hydrodynamical simulations of
turbulence in compressible gas show Kolmogorov-like 
spectra following the removal or quenching of energy sources and the 
rapid dissipation of non-eddy like motions by vorticity stretching or 
weak shocks
(Porter \etal 1992, Vazquez-Semadeni, Passot, \& Pouquet 1995).
For simulations in which energy is constantly injected or forced into 
the system, there is a continuous dissipation of energy. 
The resultant energy spectrum follows a k$^{-2}$
power law which reflects the Fourier transform of a field of velocity 
discontinuities or shocks
(Passot, Vazquez-Semadeni, \& Pouquet 1995;
Gammie \& Ostriker 1996; Padoan, Jones, \& Nordlund 1997). 
Finally, for simulations in which energy is intermittently injected 
by star formation, the resultant 
energy spectra can 
not be characterized by a singular power law as there exists simultaneous
downward and 
upward energy cascades (Vazquez-Semadeni, Passot, 
\& Pouquet 1995). 
We recognize that it is not strictly valid
to compare our results with
the results of 2D numerical simulations (in
which energy dissipation through vortex-stretching
is inhibited, and consequently energy is dissipated
solely through shocks). However it is reasonable to
suppose that k$^{-2}$ spectra are
indicative of the presence of shocks generated by
a continuous energy input.

While we measure a fairly wide distribution of $\beta$ values, the mean
of the distribution is $\sim$2.2.   From our data, we conclude that 
simulations which include 
the continuous forcing of energy into the system offer the most accurate
description of the molecular line observations.  The energy is mostly 
likely dissipated in shocks and must be replenished by some source.
Miesch \& Bally (1994) summarize various energy sources which 
may operate in the interstellar medium.  For the sample of clouds used in 
this study, the relevant sources are: outflows from
newborn stars on the smallest scales, HII regions, supernovae, stellar
winds at intermediate scales, and UV radiation, Galactic shear, gravitational
torques at the scale of the clouds.   
The lack of correspondance 
between specific values of $\beta$ and the presence or association
of HII regions and OB stars imply that these
are not the exclusive source of  energy 
in the molecular interstellar medium.  
For the sample of clouds used in this study, there are no filamentary 
signatures to 
large scale (L$>$5pc) shocks as generated 
in the forced supersonic turbulent models
of Padoan, Jones, \& Nordlund (1997).  Such shocks must occur at smaller scales
which are not resolved by these observations or are simply 
not recognized due to
the projection and blending of shock signatures. This suggests an upward 
cascade of energy powered by internal sources such as winds 
from newborn stars.

\subsection{Comparison with Results for the Atomic ISM}

Lazarian \& Pogosyan (2000, hereafter LP00) showed that 
power spectrum analysis of HI 21cm
spectral line observations can provide information
on the intrinsic velocity spectrum of neutral gas or the
intrinsic density fluctuation spectrum of neutral
gas, depending
on whether 'thin' or 'thick' velocity intervals
are examined.

Recently the formulation of LP00
was applied to HI 21cm line data from the Small 
Magellanic Cloud 
(SMC; Stanimirovic \& Lazarian 2001, hereafter SL01).
To summarize the formulation of LP00 : for a 'thin'
velocity slice (where 'thin' means that the width
of the slice in velocity space is smaller than the
turbulent velocity dispersion), the spectral slope
of the measured power spectrum of the {\it emission}
in the thin slice can be related to the intrinsic spectral
slope of the {\it velocity field}. To make comparisons
with our measurements easier, we have converted
the LP00 formalism to our energy spectrum system.
For an observed 'energy' spectrum 
of HI emission in the thin slice $E_{s}(k) \propto k^{-\beta_{s}}$,
the inferred spectral index of the intrinsic velocity
field $\beta$ (see equation 1.1) is given by :
$$ \beta = 5 - 2\beta_{s}  \eqno(4.1)$$
Hence, the expected range of 1 $\leq$ $\beta$ $\leq$ 3
corresponds to 1 $\leq$ $\beta_{s}$ $\leq$ 2.
(In the LP00 system, the corresponding
range of observed {\it power spectrum} spectral
indices, measured as {\it negative} values, ranges
between -2 and -3.)

SL01 found, for thin velocity
slices within the SMC, $\beta_{s} \sim$ 1.8 which
implies $\beta \sim$ 1.4 - ie a velocity spectral
index which is slightly less than the Kolmogorov value
($\beta$ = 5/3) and distinctly less than the
indices measured in our molecular regions.
The SMC is well-suited for analysis of this type,
since it is distant enough that divergent lines-of-sight
do not play a substantial role in the derived quantities.
However, the scales probed by SL01 ($\sim$ 30pc - 4kpc) are
not usefully compared to our molecular region data 
here (typically up to a few tens of pc). If we
assert that the large-scale results for the SMC
are applicable within the Milky Way, and that
the molecular and atomic gas are part of some
single turbulent fluid (Ballesteros-Paredes \etal 1999) then this
requires a steepening of the velocity spectrum on
scales of a few tens of pc, indicative of energy
input on smaller scales preferentially, with
either an upward energy cascade or a different energy input
mechanism operating on larger scales. However,
given the insufficient overlap between our data and
SL01, this hypothesis cannot by currently supported
by the existing data, even allowing the above
assertions to be valid. One could argue on the
basis of the same data, for example, that if the
atomic spectral slope continues similarly (as
$\sim$ Kolmogorov) to smaller scales than probed by
SL01 then molecular regions preferentially
occur in regions of enhanced energy input.
Distinguishing between these two scenarios
thus requires analysis of atomic data
on scales similar to that of the molecular
data.

Measurements of the spectral index of HI emission
in comparably 'thin' velocity slices within the Milky Way
have been made by Green (1993).
With a suitable interpretation within the LP00
formulation these  may be more directly compared to our
results. We note the caveat that these nearby
measurements, both for Green's data {\it and} ours,
may be affected by divergent line-of-sight geometry.
From spectral line data in the Outer Galaxy
(towards l,b = 140,0 - slightly below our field P1)
Green measured spectral indices of HI intensity
energy spectra ranging between 1.06 and 2.02 (representative
errors on these indices are 0.1 - 0.2). 
These imply values for the intrinsic
spectral slope of the velocity energy spectrum
covering the entire expected range of 1 $\leq$ $\beta$ $\leq$ 3.
We assume that the influence of density fluctuations
on thin slices is negligible, as it is for the SMC data.
If this is not valid, then the inferred intrinsic $\beta$
are low compared to the true values.

In the velocity range similar to our
P1 field (-50 to -30 kms$^{-1}$) the emission
energy spectrum indices
ranged between $\sim$ 1.36 to 1.75, corresponding
to intrinsic velocity field spectral indices
of $\beta \sim$ 1.5 - 2.3, while our molecular field
P1 has $\beta$ = 2.38, near the upper end
of the range for HI.
At the distance of the Perseus Arm, the range of
scales probed by Green's measurements is
similar to that of our measurements.

However, closer inspection of Green's results does
not provide any obvious interpretation of
the variation of the HI velocity index with
respect to the locations of the spiral
arms in the Outer Galaxy (defined {\it kinematically}).
This may be due in part to the complex nature of
the relationship between observed V$_{LSR}$ and 
actual physical location, due to the influence
of non-circular motions induced by the
Perseus Spiral Arm 
(Roberts 1971, Heyer \& Terebey 1998).
Since the observed index $\beta_{s}$ varies between
1.06 at V$_{LSR}$ = -63 kms$^{-1}$ and
1.75 at V$_{LSR}$ = -48 kms$^{-1}$,
the interpretation of the information provided
by these measurements 
presents a tremendous challenge.

Clearly, a more complete analysis of directly
comparable atomic and molecular regions would
be of great interest, preferably with the
same analysis methods. The Canadian Galactic
Plane Survey (English \etal 1998)
can provide such a database. Of particular
interest would be the analysis of HI self absorption
regions (Gibson \etal 2000), which trace
cooler atomic gas both with and without a detectable
molecular component and may thus probe the
supposed transition region between Kolmogorov and
shock-dominated turbulence.

\section{Conclusions}

We have identified macroscopic
velocity field correlations within an ensemble
of molecular clouds in the outer Galaxy using 
Principal Component Analysis and the characterization of
velocity and spatial scales from the eigenvectors and eigenimages.
\begin{enumerate}
\item
The velocity correlations 
are 
well-characterized by a relationship of the
form 
$$ {\delta}v \sim L^{\alpha} $$
The mean value of the spectral index from a sample of 
23 clouds is 0.62.  Using the calibration of Brunt \& Heyer (2001),
this corresponds to a spectral index to the turbulent 
energy spectrum, $\beta$=2.17.
\item These measurements of $\beta$ are consistent with 
hydrodynamic simulations of the interstellar medium in which 
there is  a continuous injection of energy into the system.
\item The absence of large scale shocks within the molecular 
clouds imply that the energy is injected at scales $<$ several pc.
\end{enumerate}

\acknowledgments

We thank E. Vazquez-Semadeni, J. Stone, E. Ostriker, J. Scalo,
E. Falgarone for many useful comments over the course of this project.
This work was supported by NSF grant AST 97-25951  to the Five College
Radio Astronomy Observatory. 
The Dominion Radio Astrophysical 
Observatory is a National Facility 
operated by the National Research 
Council. The Canadian Galactic Plane 
Survey is a Canadian project with 
international partners, and is supported 
by the Natural Sciences and Engineering 
Research Council (NSERC).

\clearpage
\appendix
\section {Appendix}
\subsection{Removing the Noise Contribution to the ACF}

In order to derive accurate spatial scales independent 
of the signal to noise level of the data, 
it is necessary to remove the noise contribution to the
eigenimage ACFs.
For Gaussian noise,
there is no contribution to an ACF for lags
$\tau \neq 0$. The contribution at zero lag is simply
equal to the variance of the noise ($\sigma^{2}_{N}$).
Consequently subtracting this quantity from the
``raw'' eigenimage ACF at lag zero, prior to normalization, will
remove the noise contribution, with residual standard deviation 
of $\sigma^{2}_{N}/\sqrt{N_{p}}$
where $N_{p}$ is the number of pixels used to calculate
the ACF. Very accurate estimates
of $\sigma^{2}_{N}$ can be obtained from several high $l$
eigenimages, due to the propagation of noise to the eigenimages
(Heyer \& Schloerb 1997). 

For the Outer Galaxy Survey data, the noise properties are
distinguished from gaussian noise due to the reference
sharing technique used to take the data (Heyer \etal 1998).  
There are 
noise correlations for $\tau \neq 0$.
An ACF of the noise of the Outer Galaxy Survey
contains correlations at
($|\tau_{y}| = 0, 1\;$ pixels $\;$ and $\; 
|\tau_{x}| = 0, 5, 10, 15, 20, 25\;$ pixels)
where $x$ is along the Galactic longitude direction and $y$ along the
Galactic latitude direction.
The correlation at non-zero $\tau_{x}$ is due
the 5 pixel dimension of the focal plane array receiver and the 
sharing of a single reference measurement over 5 settings of the array
in the x direction.
The correlation at non-zero $\tau_{y}$ is due to sharing the reference
measurement 
over 2 settings of the array in the y direction.

We have developed a simple method to remove the correlated noise 
contribution to the eigenimage ACF. 
Given the eigenimage, 
$$ I^{l}({\bf r}) \; = \; I^{l}_{0}({\bf r}) + N({\bf r}) $$
where I$^{l}_{0}({\bf r})$ is a noiseless eigenimage
and N(${\bf r}$) is the noise contribution, the
unnormalized spatial,
C$^{l}_{I}(\mbox{\boldmath $\tau$}$), is 
$$ C^{l}_{I}(\mbox{\boldmath $\tau$}) \; = \; < (I^{l}_{0}({\bf r})+N({\bf r}))(I^{l}_{0}({\bf r} + \mbox{\boldmath $\tau$})+N({\bf r} + \mbox{\boldmath $\tau$})) > $$
$$ C^{l}_{I}(\mbox{\boldmath $\tau$}) \; = \; C^{l}_{I0}(\mbox{\boldmath $\tau$}) + C^{l}_{N}(\mbox{\boldmath $\tau$}) + < I^{l}_{0}({\bf r})N({\bf r} + \mbox{\boldmath $\tau$}) + I^{l}_{0}({\bf r} + \mbox{\boldmath $\tau$})N({\bf r}) > $$
where C$^{l}_{I0}$ is the ACF of the noiseless eigenimage
and C$^{l}_{U}$ is the ACF of the noise.
Assuming the last of these terms (the correlation of signal and
noise) is zero, then 
$$ C^{l}_{I}(\mbox{\boldmath $\tau$}) \; = \; C^{l}_{I0}(\mbox{\boldmath $\tau$}) + C^{l}_{N}(\mbox{\boldmath $\tau$}) $$
and we may approximate the noiseless eigenimage ACF by subtracting
an estimate of a pure noise ACF from the raw eigenimage ACF. Pure
noise ACFs can be obtained from several high $l$ eigenimages that
are free of signal. We refer to this process as ``ACF Flat Fielding''
since it is similar to the process of removing sensitivity variations
across a CCD chip.
An example of the process is shown in Figure~10. This
is the fifth eigenimage from field L1 
in raw form (C$_{I}$)
and in flat-fielded form (C$_{IO}$) after subtraction of the
flat field (C$_{N}$). These are 1D cuts through the ACFs showing
the noise correlations along the x and y axes. After
flat-fielding, this ACF (as in the majority of cases) is
very isotropic within the e-folding length.
\clearpage
\begin{figure}
\plotone{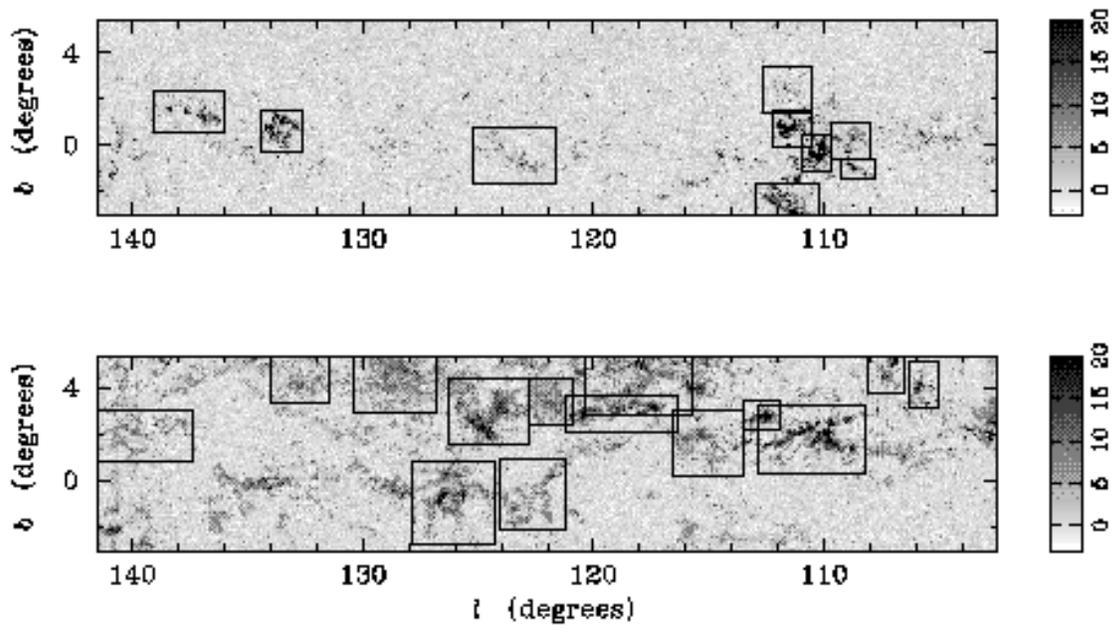}
\caption
{Images of $^{12}$CO J=1-0 intensity integrated over the velocity 
intervals of the Perseus (top) and local (bottom) spiral arms.  The 
boxes show the individual fields analyzed in this study.  
}
\end{figure}

\begin{figure}
\plotone{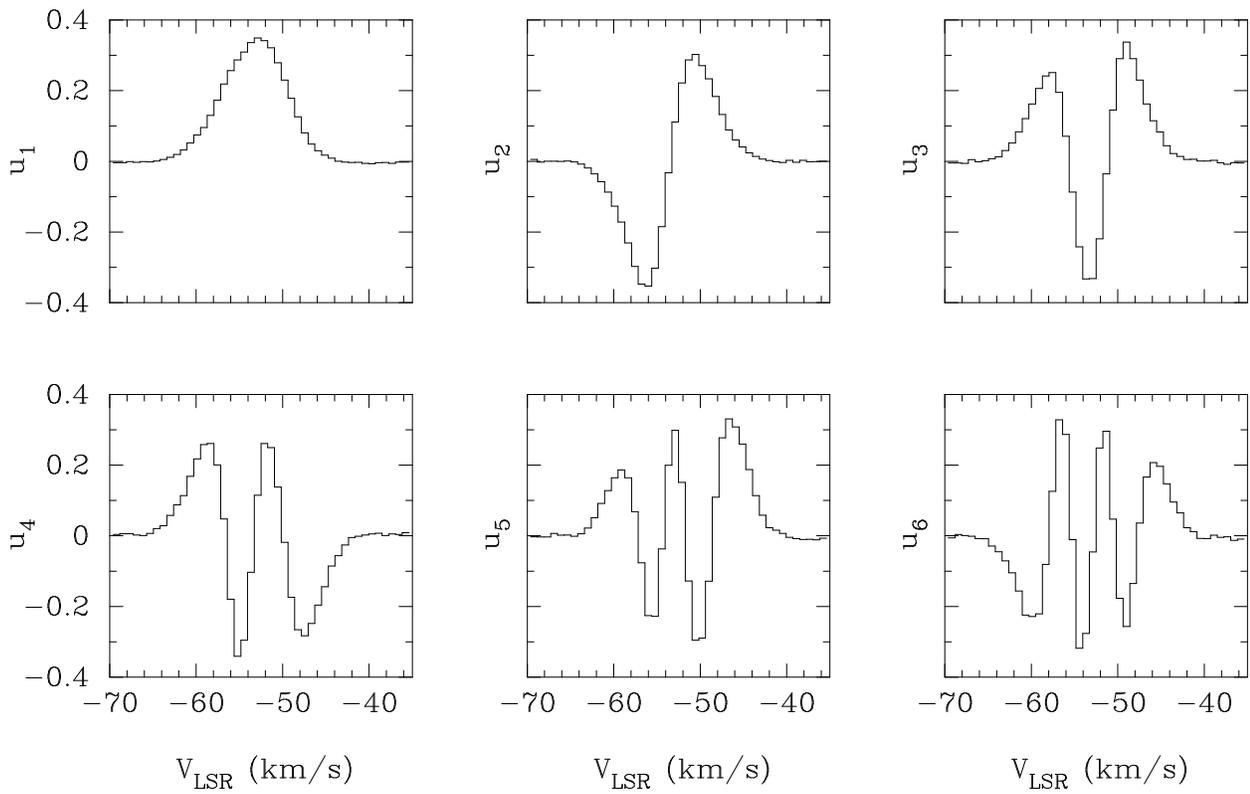}
\caption
{a) The set of eigenvectors, $u_l(v)$, ($ 1 \le l \ge 6$) 
derived for the P6 (NGC~7538) field. }
\end{figure}

\begin{figure}
\plotone{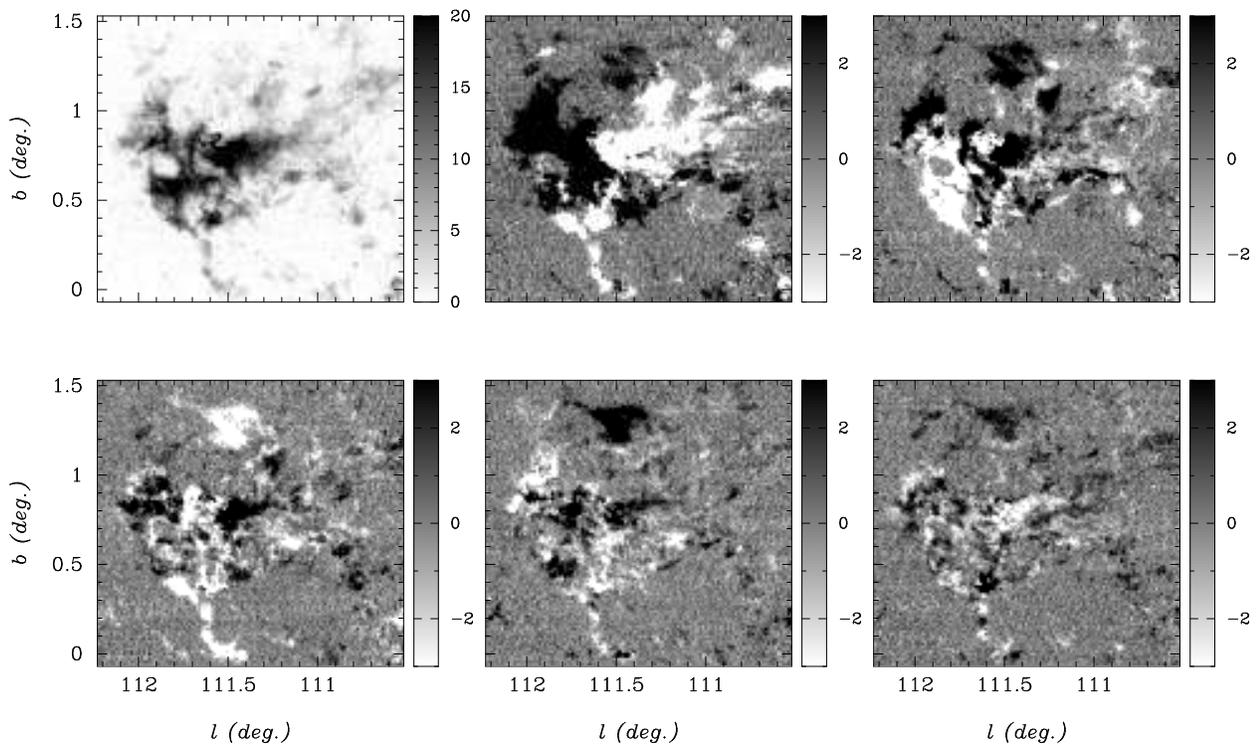}
\caption {
b) The set of eigenimages, $I_l(r)$, derived for the P6 field from the 
projection of the spectra onto the eigenvectors.}
\end{figure}

\begin{figure}
\plotone{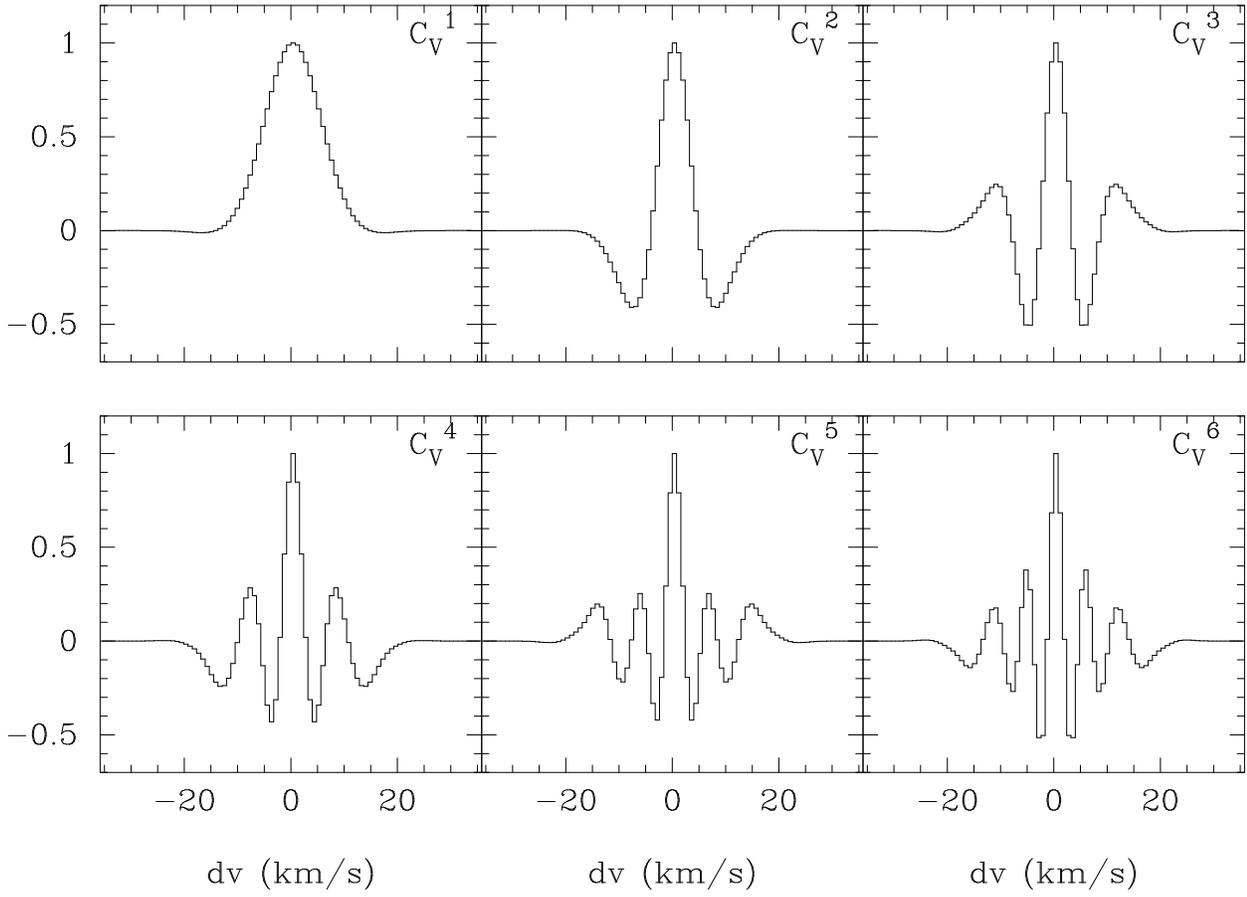}
\caption
{a) The 1D autocorrelation functions, $C^l_V(dv)$ of the eigenvectors for 
the P6 field from which the characteristic scale, ${\delta}v_l$
is determined from the e-folding length.}
\end{figure}

\begin{figure}
\plotone{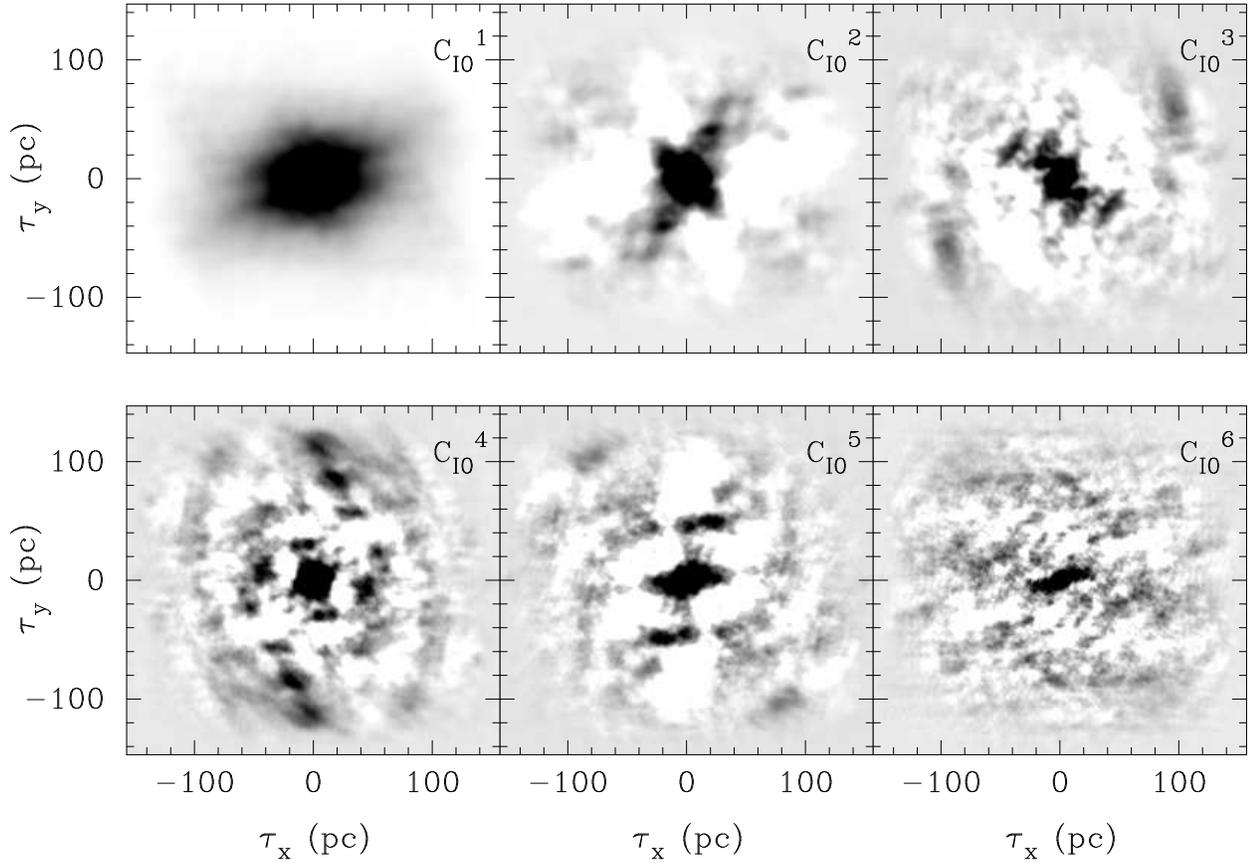}
\caption
{b) The 2D autocorrelation functions, $C^l_{I0}(\tau_x,\tau_y)$ of the eigenimages
of 
the P6 field. The characteristic spatial scale, $L_l$, is derived for each
principal component, $l$, from the e-folding lengths of 
$C^l_{I0}(\tau_x,\tau_y)$ respectively along the cardinal directions.}
\end{figure}

\clearpage

\begin{figure}
\plotone{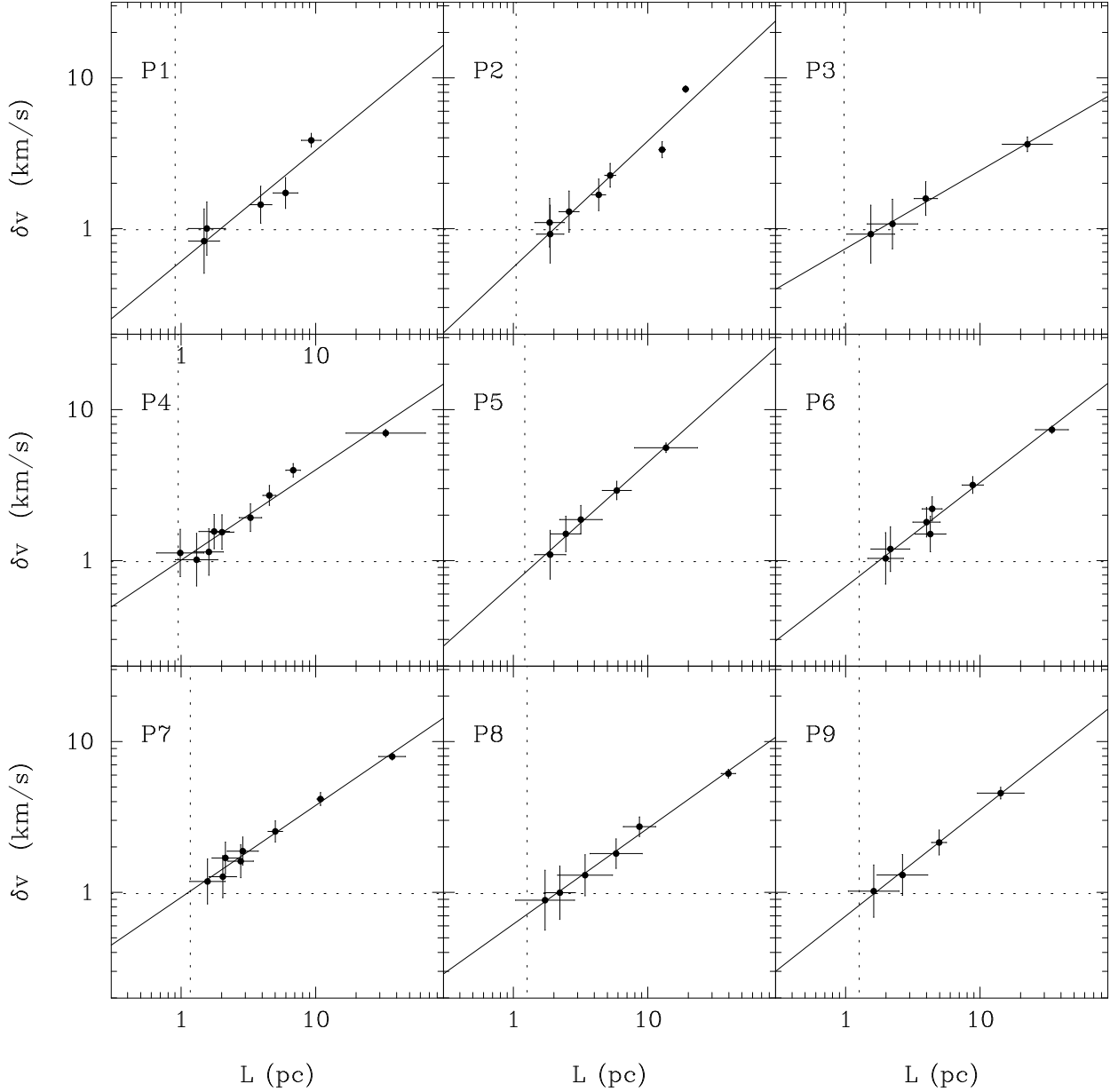}
\caption
{The derived velocity scales versus spatial scales for all
significant components for the Perseus arm fields.  The solid 
line shows a power law fit to the ${\delta}v$,L points.  The 
spectral index and amplitude for each source are reported in 
Table~2.  The
horizontal and vertical dotted lines show the spectral and spatial 
 resolution limits of the data.  Points with ${\delta}v_l$ greater than the 
spectral resolution and $L_l$ less than  twice the spatial resolution are 
excluded from the fits. }
\end{figure}

\begin{figure}
\plotone{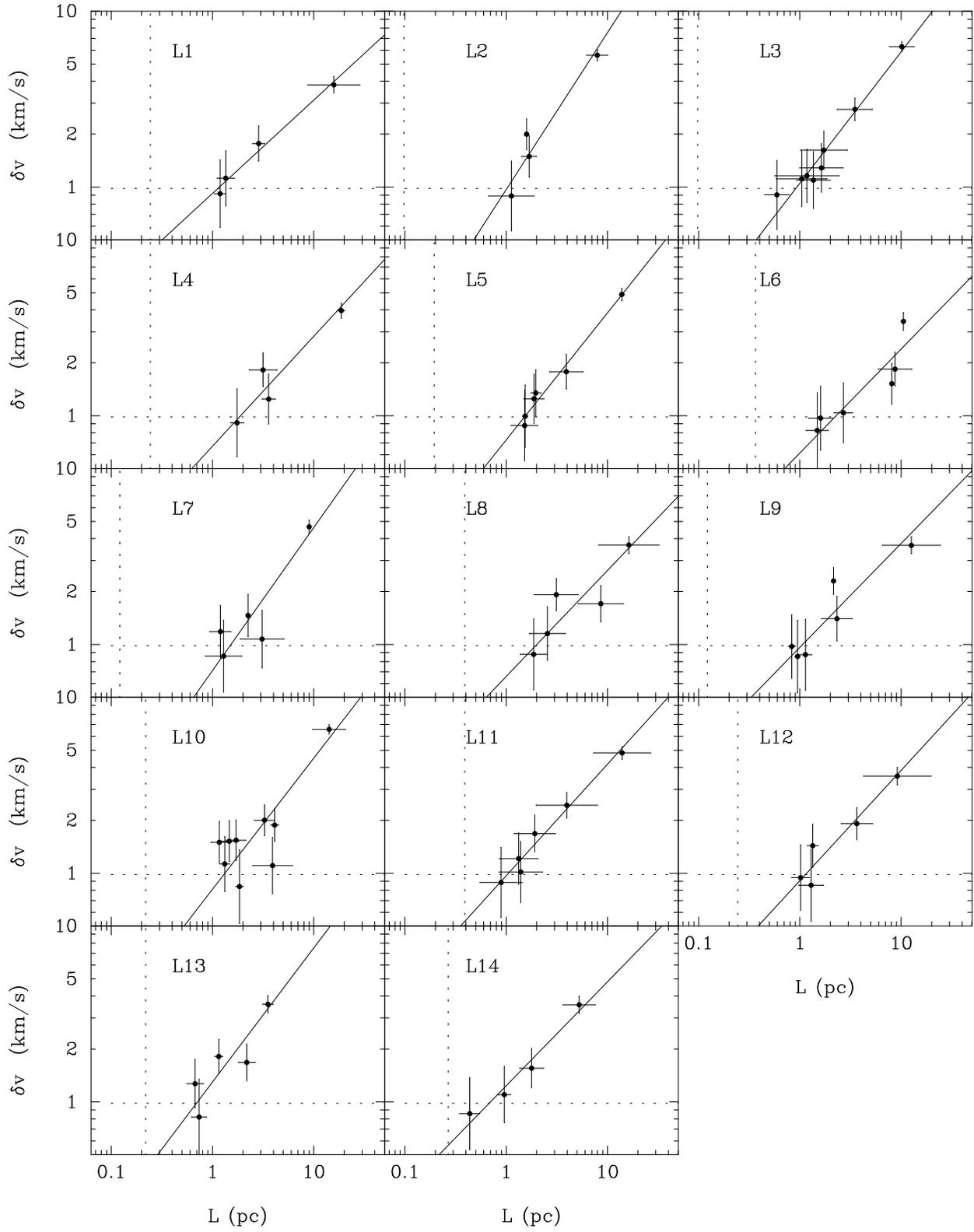}
\caption {Same as Figure 4 for the local spiral arm sources L1-L14.}
\end{figure}

\clearpage

\begin{figure}
\plotone{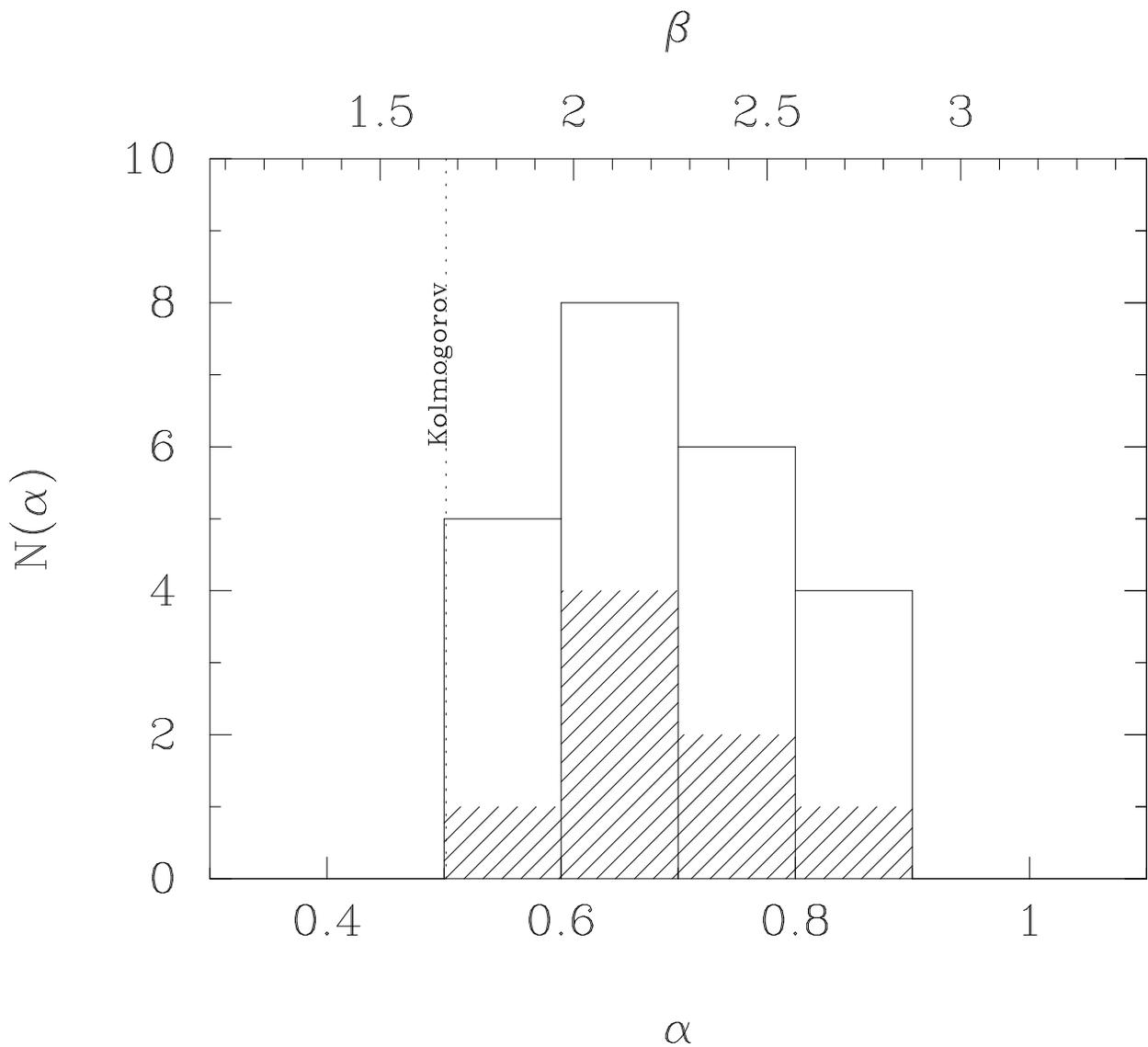}
\caption
{Distribution of measured values of $\alpha$ for all selected 
fields. The corresponding values of $\beta$ using the
calibration of Brunt \& Heyer (2001) are shown as the top 
ordinate axis.  The hatched areas in the histogram denote the
contributions of those clouds associated with HII regions and OB stars.
The weighted ($1/\sigma_\alpha^2$) mean  value
for $\beta$ is  2.17$\pm$0.3.  The vertical dotted line corresponds to 
$\beta=5/3$ for Kolmogorov turbulence.}
\end{figure}

\begin{figure}
\plotone{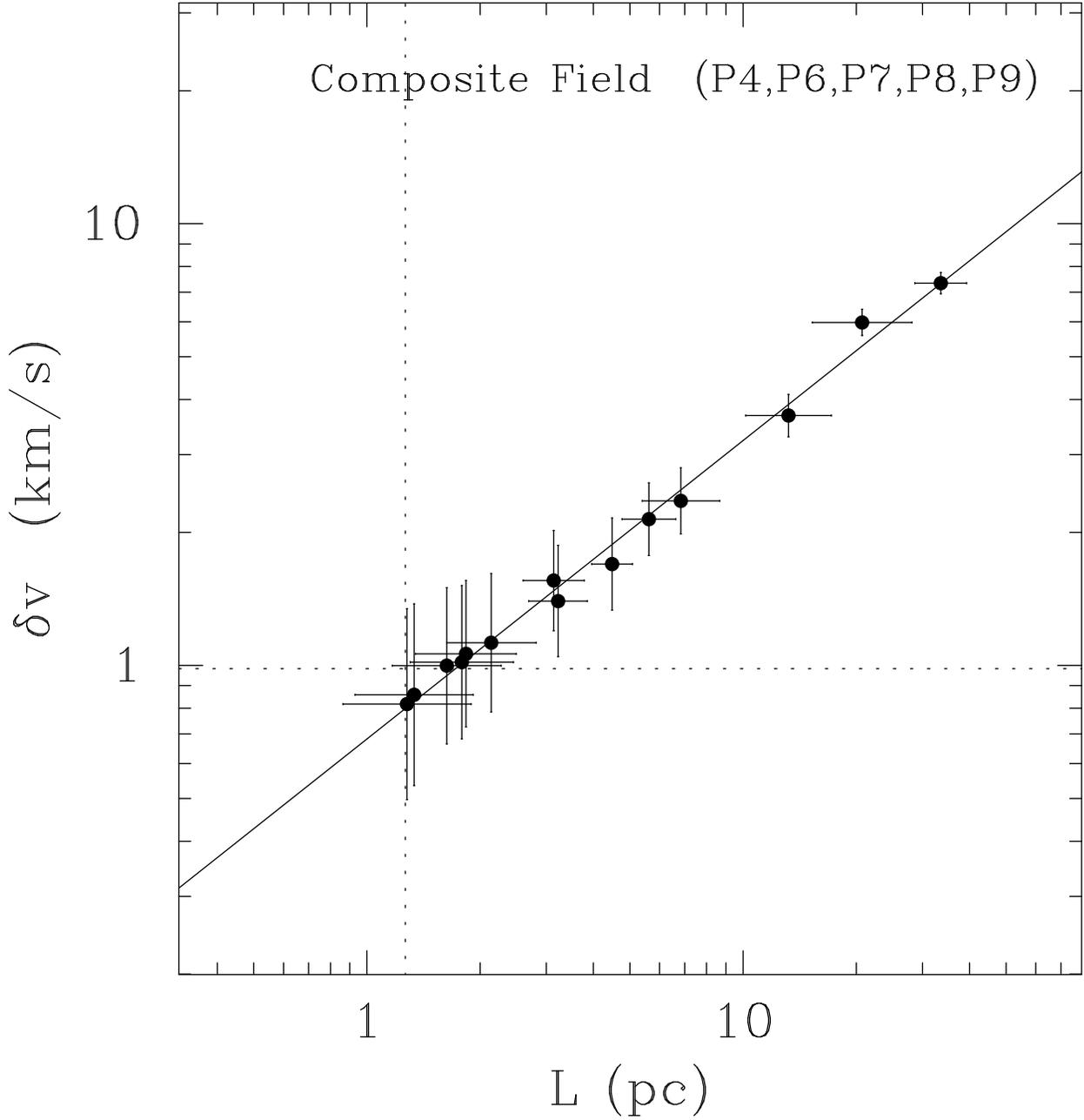}
\caption
{Results from the analysis upon a larger field in the Perseus arm
comprised of fields P4,P6,P7,P8, and P9.  The spectral index of the
composite field is similar to the geometric mean value of the 
four subfields.}
\end{figure}

\begin{figure}
\plotone{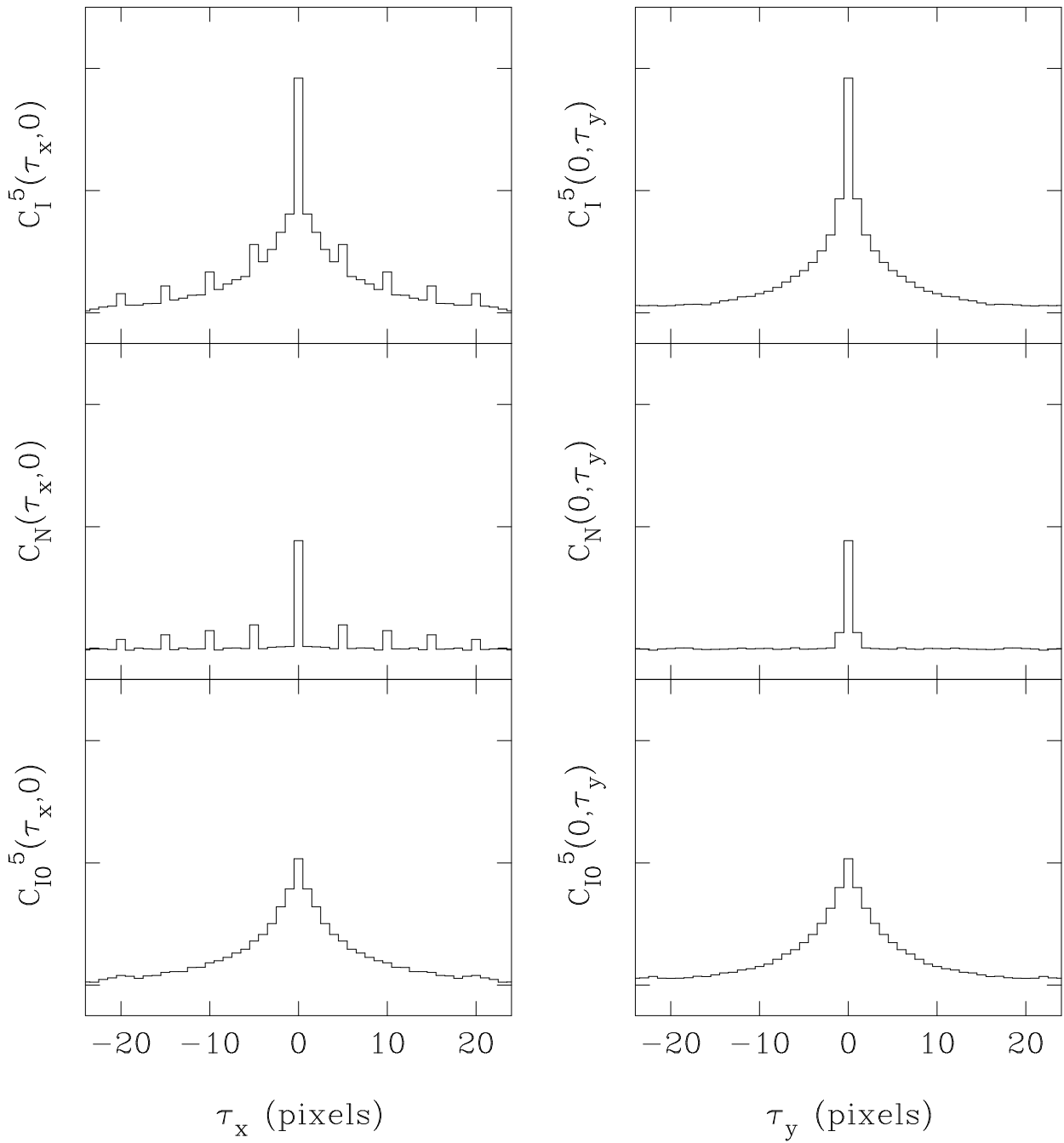}
\caption
{A demonstration of correcting for the correlated noise properties
in the autocorrelation function.
(top) the uncorrected ACFs, $C^5_{I}(\tau_x,0), C^5_{I}(0,\tau_y)$ along the 
x and y directions respectively.   
(middle) correlated noise ACFs $C_{N}(\tau_x,0), C_{N}(0,\tau_y)$ 
derived from the last stored eigenimage ($l=32$).
(bottom) corrected ACFs, $C^5_{I0}(\tau_x,0), C^5_{I0}(0,\tau_y)$ 
after the subtraction of the noise ACFs from the raw, uncorrected ACFs.}
\end{figure}

\begin{table}
\begin{center}
\caption
{Boundaries of the selected
Fields}
\vspace{7mm}
 \begin{tabular}
{cccccccccccc}
\hline
Field & $v_{min}$ & $v_{max}$ & $l_{max}$ & $l_{min}$ & $b_{min}$ & $b_{max}$ & N$_{v}$ & N$_{l}$ & N$_{b}$ & Distance (kpc) & Comments\\
 \hline
P1 & -52.0 & -30.1 & 139.09 & 135.97 & 0.54 & 2.40 & 28 & 220 & 134 & 3.7 & W5\\
P2 & -60.2 & -30.1 & 134.43 & 132.65 & -0.24 & 1.53 & 38 & 128 & 128 &4.3 & W3\\
P3 & -56.1 & -32.5 & 125.22 & 121.61 & -1.63 & 0.75 & 30 & 260 & 172 & 4.0 \\
P4 & -55.3 & -25.2 & 112.94 & 110.25 & -3.03 & -1.70 & 38 & 194 & 96 & 3.9 \\
P5 & -64.2 & -39.0 & 112.61 & 110.59 & 1.44 & 3.46 & 32 & 146 & 146 & 5.0 \\
P6 & -69.9 & -35.0 & 112.23 & 110.52 & -0.07 & 1.53 & 44 & 124 & 116 & 5.2 & NGC~7538\\
P7 & -69.9 & -30.0 & 110.98 & 109.68 & -1.12 & 0.46 & 50 & 94 & 114 & 4.8 & Sh~156 \\
P8 & -65.0 & -35.0 & 109.65 & 107.96 & -0.61 & 0.99 & 38 & 122 & 116 & 5.2\\
P9 & -61.8 & -41.5 & 109.32 & 107.80 & -1.49 & -0.64 & 26 & 110 & 62 & 5.2 & Sh~152\\
L1 & -20.3 & -3.3 & 141.54 & 137.37 & 0.88 & 3.10 & 22 & 300 & 160 & 1.0\\
L2 & -14.7 & 4.8 & 134.00 & 131.43 & 3.39 & 5.41 & 25 & 186 & 146 & 0.4 \\
L3 & -20.3 & 4.8 & 130.38 & 126.82 & 2.97 & 5.41 & 32 & 256 & 176 & 0.4 \\
L4 & -20.3 & -7.3 & 127.87 & 124.31 & -2.75 & 0.81 & 17 & 256 & 256 & 1.0 \\
L5 & -20.3 & -3.3 & 126.33 & 122.78 & 1.58 & 4.44 & 22 & 256 & 206 & 0.8 \\
L6 & -28.5 & -9.8 & 124.10 & 121.19 & -2.14 & 1.00 & 24 & 210 & 226 & 1.5\\
L7 & -9.8 & 4.8 & 122.85 & 120.91 & 2.48 & 4.42 & 19 & 140 & 140 & 0.5 \\
L8 & -25.2 & -10.6 & 121.17 & 116.30 & 2.13 & 3.65 & 19 & 350 & 110 & 1.6 \\
L9 & -9.8 & 4.8 & 120.34 & 115.72 & 2.83 & 5.41 & 19 & 332 & 186 & 0.5 \\
L10 & -25.2 & 4.8 & 116.57 & 113.51 & 0.18 & 3.10 & 38 & 220 & 210 & 0.9 \\
L11 & -26.8 & -9.8 & 113.50 & 111.95 & 2.19 & 3.51 & 22 & 112 & 96 & 1.6 \\
L12 & -17.9 & 0.8 & 112.92 & 108.21 & 0.32 & 3.24 & 24 & 338 & 210 & 1.0 &Cep~OB3\\
L13 & -20.3 & 4.8 & 108.12 & 106.57 & 3.84 & 5.41 & 32 & 112 & 114 & 0.9 &Sh~140\\
L14 & -20.3 & 4.8 & 106.36 & 105.01 & 3.17 & 5.19 & 32 & 98 & 146 &  1.1 \\
\hline    
 \end{tabular}
\label{fields}
\end{center}
\end{table}

\begin{table}
\begin{center}
\caption{PCA Results }
\vspace{7mm}
 \begin{tabular}
{cccccccc}
\hline
Field & $\zeta_{1}$ & N$_{pc}$ & N$_{c}$ & $\alpha$ & c & $\beta$ & $\gamma$\\
\hline
P1 & 4.2 & 5 & 0 & 0.73$\pm$0.10 & 0.61$\pm$0.07 & 2.38$\pm$0.38 & 0.66$\pm$0.11 \\
P2 & 7.6 & 7 & 0 & 0.84$\pm$0.09 & 0.56$\pm$0.06 & 2.69$\pm$0.36 & 0.75$\pm$0.10 \\
P3 & 2.5 & 4 & 0 & 0.52$\pm$0.01 & 0.74$\pm$0.02 & 1.72$\pm$0.24 & 0.33$\pm$0.02 \\
P4 & 5.7 & 9 & 0 & 0.60$\pm$0.04 & 1.01$\pm$0.06 & 1.97$\pm$0.28 & 0.47$\pm$0.07 \\
P5 & 1.8 & 5 & 1 & 0.80$\pm$0.02 & 0.71$\pm$0.04 & 2.58$\pm$0.25 & 0.72$\pm$0.06 \\
P6 & 11.7 & 7 & 0 & 0.69$\pm$0.02 & 0.67$\pm$0.04 & 2.25$\pm$0.25 & 0.62$\pm$0.06 \\
P7 & 11.1 & 8 & 0 & 0.61$\pm$0.02 & 0.93$\pm$0.05 & 2.00$\pm$0.25 & 0.49$\pm$0.04 \\
P8 & 5.0 & 6 & 0 & 0.63$\pm$0.01 & 0.62$\pm$0.01 & 2.08$\pm$0.25 & 0.53$\pm$0.03 \\
P9 & 6.7 & 4 & 0 & 0.70$\pm$0.02 & 0.70$\pm$0.03 & 2.28$\pm$0.25 & 0.63$\pm$0.06 \\
L1 & 3.2 & 4 & 1 & 0.53$\pm$0.02 & 0.92$\pm$0.05 & 1.76$\pm$0.25 & 0.36$\pm$0.04 \\
L2 & 3.8 & 4 & 0 & 0.89$\pm$0.07 & 0.97$\pm$0.13 & 2.86$\pm$0.33 & 0.81$\pm$0.09 \\
L3 & 4.8 & 8 & 0 & 0.75$\pm$0.04 & 1.05$\pm$0.07 & 2.42$\pm$0.28 & 0.67$\pm$0.07 \\
L4 & 4.5 & 4 & 0 & 0.62$\pm$0.02 & 0.67$\pm$0.08 & 2.04$\pm$0.25 & 0.52$\pm$0.05 \\
L5 & 5.2 & 6 & 0 & 0.72$\pm$0.02 & 0.73$\pm$0.04 & 2.34$\pm$0.25 & 0.65$\pm$0.06 \\
L6 & 3.4 & 6 & 0 & 0.59$\pm$0.10 & 0.62$\pm$0.08 & 1.94$\pm$0.39 & 0.46$\pm$0.17 \\
L7 & 5.8 & 5 & 0 & 0.81$\pm$0.07 & 0.71$\pm$0.13 & 2.61$\pm$0.32 & 0.73$\pm$0.09 \\
L8 & 5.0 & 5 & 0 & 0.60$\pm$0.04 & 0.66$\pm$0.07 & 1.98$\pm$0.27 & 0.48$\pm$0.07 \\
L9 & 4.4 & 6 & 0 & 0.59$\pm$0.05 & 0.96$\pm$0.07 & 1.94$\pm$0.29 & 0.46$\pm$0.09 \\
L10 & 3.4 & 9 & 0 & 0.74$\pm$0.05 & 0.81$\pm$0.15 & 2.41$\pm$0.28 & 0.67$\pm$0.07 \\
L11 & 6.0 & 6 & 0 & 0.63$\pm$0.03 & 0.97$\pm$0.05 & 2.07$\pm$0.25 & 0.53$\pm$0.05 \\
L12 & 5.8 & 5 & 1 & 0.63$\pm$0.05 & 0.91$\pm$0.10 & 2.05$\pm$0.28 & 0.52$\pm$0.08 \\
L13 & 8.0 & 5 & 1 & 0.76$\pm$0.11 & 1.30$\pm$0.16 & 2.46$\pm$0.41 & 0.68$\pm$0.12 \\
L14 & 7.9 & 4 & 0 & 0.59$\pm$0.05 & 1.23$\pm$0.07 & 1.94$\pm$0.28 & 0.46$\pm$0.08 \\
\hline
 \end{tabular}
\label{alpharesults}
\end{center}
\end{table}

\end{document}